\documentclass{llncs}

\usepackage[utf8]{inputenc}
\usepackage[english]{babel}
\usepackage{amssymb,amsmath,textcomp}
\usepackage{jmllistings}
\usepackage{mathtools}
\usepackage{booktabs}
\usepackage{listings}
\usepackage{wrapfig}
\usepackage{tikz}
\usepackage{url}
\usepackage{xspace}
\usepackage{tikz}
\usepackage{paralist}
\usepackage{pifont}
\usetikzlibrary{arrows}
\usetikzlibrary{positioning,calc,fit}

\usepackage{courier}

\lstdefinelanguage{KPS}{%
  keywords={%
    script,foreach,theonly,case,cases,
    next,init,boolean,unsigned,signed,word,INIT,TRANS
  },
  morecomment=[l]{//},
  sensitive=true,
  breaklines=true,
  extendedchars=true
}

\usepackage[inline]{enumitem}
\usetikzlibrary{arrows}
\usepackage[]{algorithm2e}

\newcommand{\KeY}{Ke\kern-0.1emY\xspace}
\newcommand{\xparagraph}[1]{\textbf{#1}}
\newcommand{\lstt}[1]{\lstinline[basicstyle=\tt,
keywordstyle=\tt, showspaces=false, showstringspaces=false,]{#1}}
\newcommand\keyword[1]{\texttt{#1}} 
\newcommand\kwTheOnly{\keyword{theonly}\xspace}
\newcommand\kwForeach{\keyword{foreach}\xspace}
\newcommand\kwCases{\keyword{cases}\xspace}
\newcommand\kwCase{\keyword{case}\xspace}
\newcommand\kwRepeat{\keyword{repeat}\xspace}
\newcommand\kwWhile{\keyword{while}\xspace}
\newcommand\kwIf{\keyword{if}\xspace}
\newcommand\kwDefault{\keyword{default}\xspace}
\newcommand{\PSDBG}{\mbox{\textsc{psdbg}}\xspace}
\newcommand{\KPS}{\textsc{kps}\xspace}

\newcommand\myparagraph[1]{\xparagraph{#1.}}

\author{Bernhard Beckert \and
Sarah Grebing \and Alexander Weigl}
\institute{Karlsruhe Institute of Technology \\
\url{{beckert,sarah.grebing,weigl}@kit.edu}}
\title{Debugging Program Verification Proof Scripts \\[-\medskipamount] {\scriptsize -- Tool
  Paper --}}

\usepackage{comment}

\renewcommand\emph[1]{{\textit{#1}}}

\begin{document}

\maketitle

\begin{abstract}
  Interactive program verification is characterized by iterations of unfinished
  proof attempts. To support the process of constructing a complete proof, many
  interactive program verification systems offer a proof scripting language as
  a text-based way to describe the non-automatic steps in a proof. Such
  scripting languages are beneficial, but users spent a lot of effort on
  inspecting proof scripts and the proofs they construct to detect the cause
  when a proof attempt is unsuccessful and leads to unintended proof states.
  We present an offline and replay debugger to support the user in analyzing
  proof attempts performed with proof scripts. This debugger adapts successful
  concepts from software debugging to the area of proof script debugging.
  The tool is built on top of \KeY{}, a system for deductive verification of
  Java programs. The debugger and its graphical user interface are designed to
  support program verification in particular, the underlying concepts and the
  implementation, however, are adaptable to other provers and proof tasks.
\end{abstract}

\section{Introduction}

\xparagraph{Motivation.}
Proving complex properties of programs requires user guidance, which can come in the form
of program annotations as well as
user interaction during proof construction. Providing the right guiding information that
allows a verification system to find a proof is, in general, an iterative
process of repeated failed attempts.
Also, the characteristics of program verification proofs are considerably different
from proofs of mathematical theorems (such as properties of algebraic structures).
%
Proofs in program verification consist of many structurally
and/or semantically similar cases that are syntactically large, but usually of
low intrinsic complexity.
The mechanism for providing user guidance needs to reflect this peculiarity of
proofs in the program verification domain and provide appropriate means for
interaction. To support the iterative process of constructing proofs, many
interactive program verification systems offer a proof scripting language as
a text-based way to describe the non-automatic steps in a proof. Such scripting
languages are beneficial, but users spent a lot of effort on inspecting proof
scripts and the proofs they construct to detect the cause when a proof attempt
is unsuccessful and leads to unintended proof states.

\xparagraph{Contribution.} In this paper, we describe our tool \PSDBG{},
an offline and
replay debugger adapting successful concepts from software
debugging to the area of proof script debugging. It implements
 our interaction concept
for interactive program verification described in~\cite{hvc2017}. \PSDBG{}
combines point-and-click with text-based interaction based on
a scripting language for proofs, \KPS{} (Sect.~\ref{sec:language}). The replay
functionalities offered by our tool allow the user to analyze unfinished proof
attempts by using functionalities known from software debugging
like forward stepping, breakpoints, and the visualization of the proof
script state and proof. Furthermore, step-back and replay of proof commands are
supported.
Partial proof scripts can be extended by either appending new script
commands (text-based) or by point-and-click selection of proof rules and
commands.
Additionally, \PSDBG contains aids for writing proof scripts such as the
generation of a case-distinction expressions for goal selection
and a visualization of the result of term matching
expressions.

\PSDBG is available at~\url{formal.iti.kit.edu/psdbg} together with
a video of its usage.
%
%
%
%

\xparagraph{Underlying verification system.}
We have chosen to built \PSDBG{} on top of the \KeY{} system, which is
an interactive theorem prover for the verification of Java programs at source
code level~\cite{KeYBook2}. \KeY{} is based on a sequent calculus for Java
Dynamic Logic. It
 was successfully applied to verify real world Java programs, e.g.,
implementations of Timsort \cite{DBLP:conf/cav/GouwRBBH15} and Dual-Pivot
Quicksort~\cite{DBLP:conf/vstte/BeckertSSU17}. 
\KeY{} constructs an explicit proof object, i.e., all proof steps and rule applications are
available to the user at any time in addition to
the current open goals.
This enables a more fine-grained stepping functionality down to the level of
single calculus rule applications.
\KeY{} offers point-and-click interaction for prover guidance. Combining \KeY{}
with a script component allows to automate user actions without losing the
advantages of point-and-click interaction. Also, this combination provides more stable
guidance in situations where the proof problem evolves.
\PSDBG is designed to support program verification in particular, the underlying concepts and the
implementation, however, are adaptable to other provers and proof tasks (Sect.~\ref{sec:architecture}).

\xparagraph{Related work.}
The need to analyze failed proof attempts in interactive theorem proving
has lead to different mechanisms for gaining insight into proof construction.
The interactive theorem provers
Isabelle~\cite{Isabelle} and Coq~\cite{Coqbook}
both provide text-based
interaction, and the way in which proofs are constructed allows
to step over tactics, to revert a tactic application, and to add tactic invocations iteratively.
The user can inspect the proof states between tactic applications. To get
a deeper insight into tactics, both tools allow for the use of debuggers for
the language in which the tactics and the tools are implemented (Standard ML
respectively OCaml). While tactics implement generic proof strategies
independent from the concrete proof problem, proof scripts are usually
tailored to the current verification task.
This difference manifests itself when debugging proof scripts in contrast to debugging tactics.
%
Additionally, Hupel proposes an interactive tracing of Isabelle's simplification
tactic~\cite{Hupel14}.
\textsc{Lean}'s metaframework~\cite{metaframework} -- an API to the theorem prover
\textsc{lean} -- provides support for classical program debuggers to
step through the execution of the declarative language of Lean.

%

%
%


\section{Language for Proof Scripts}
\label{sec:language}

In this section, we introduce the basic concepts of the \textsc{Key Proof
  Script}~(\KPS) language. As an example, we use a script constructing a proof
for the correctness of the pivotal split in a Quicksort implementation (see
Fig.~\ref{fig:exampleqs}).\footnote{The full Java source code and its
  specification can be found in
  Appendix~\ref{sec:quicksortJava}.}

\begin{figure}[tb]
\begin{lstlisting}[language=KPS]
script quicksort_split() {
  autopilot_prep;       //perform symbolic execution and simplify
  foreach { tryclose; } //try to close all trivial cases
  foreach { simp_upd;   seqPermFromSwap;  andRight; }
  cases {
    case match `==> seqDef(_,_,_) = seqDef(_,_,_)`: auto;
    case match `==> (\exists ?X (\exists ?Y _))` :
        instantiate  var=X with=`i_0`;
        instantiate  var=Y with=`j_0`;
        auto;
  } }
\end{lstlisting}
  \caption{A proof script for proving correctness of the split method of
    Quicksort (see Appendix~\ref{sec:quicksortJava}, line 76). The first lines
    perform a pre-processing. After application of simplification steps and
    a rule specific for the data type sequence (\texttt{seqPermFromSwap}), user
    guidance in the form of quantifier instantiations is required (lines 8--9).
    The match expression in line~7 matches sequents that contain a formula which
    consists of at least two nested existential quantifiers and binds the
    concrete terms of the quantified variables to the schema variables
    \texttt{?X} resp.~\texttt{?Y} to be used in in lines~8 and 9 where they
    are parameters for the proof command \texttt{instantiate}.
  }
  \label{fig:exampleqs}
\end{figure}

%
%
%
%
%
%
%
%
%
%
%

\myparagraph{State} A \emph{proof state} consists of a set of \emph{proof goals}
of which at most one is \emph{selected}.
The main part of a proof goal is an open verification condition. In addition,
it assigns values to variables.
These variables are \emph{goal-local}, i.e., changing the value of a variable
has only local effect. When a new goal is created, it inherits its parent
goal's assignment.
The configuration of the underlying theorem prover
(e.g., the particular heuristic used for proof search)
is accessible and can be changed via a special subset of these variables.
%
%
%

\myparagraph{Mutators}
As proof construction is characterized by selecting and manipulating goals, \KPS
provides \emph{goal selectors} and \emph{mutators}.
Mutators are commands which modify a single proof goal by either manipulating
the verification condition or changing the variable assignment.
Mutators for verification conditions are either calls to sub-scripts (to
construct sub-proofs) or
commands from the underlying theorem prover. In the example in
Fig.~\ref{fig:exampleqs}, one of the mutators is \texttt{autopilot\_prep} (line~2), an
internal prover strategy of \KeY{}, that performs symbolic execution of the
program to be verified with intermediate simplification steps. Another mutator in
the example is \texttt{instantiate} (lines~8 and~9), which is a rule
with parameters \texttt{var} and \texttt{with}. This rule instantiates the
quantified variable \texttt{var} with a term.
%
%
If a proof state contains more than one goal, before applying mutators a single
goal has to be selected, as described in the following.

\myparagraph{Goal selectors} With goal selectors one picks goals from
the current state for mutator application.
%
\KPS provides the following selectors: \kwForeach, \kwTheOnly, \kwCases. With
\kwForeach, a mutator is applied to all proof goals (lines~3 and~4 in Fig.~\ref{fig:exampleqs}).
The \kwCases selector is used to make case distinctions over proof goals based
on matching pattern expressions (in lines~5 to 10 there are two cases).
In addition to syntactical patterns, matching expressions can be semantic, and
they can refer to
local goal variables; see~\cite{hvc2017} for more details.
%
This kind of selection statement allows to mutate similar goals in
the same way.
%
After the evaluation of
a matching expression, the state is updated
by  variable bindings.

\section{Tool Features and Their Use}
Program verification is an iterative process of unsuccessful proof
attempts.
The user needs to find the reason why proof construction failed.
%
In the case of proof debugging, the user investigates whether the last state of the
proof script, including the remaining verification conditions, matches his or her
mental model of the proof, and how this state was reached.
%
%
%
%
%
To support the user, our tool makes use of the analogy between writing programs
and writing proof scripts presented in~\cite{hvc2017}.
This analogy enables us to adopt mechanisms from software debugging systems to
the analysis of failed proof attempts.

\begin{figure}[h!]
  \centering
  \includegraphics[width = \textwidth]{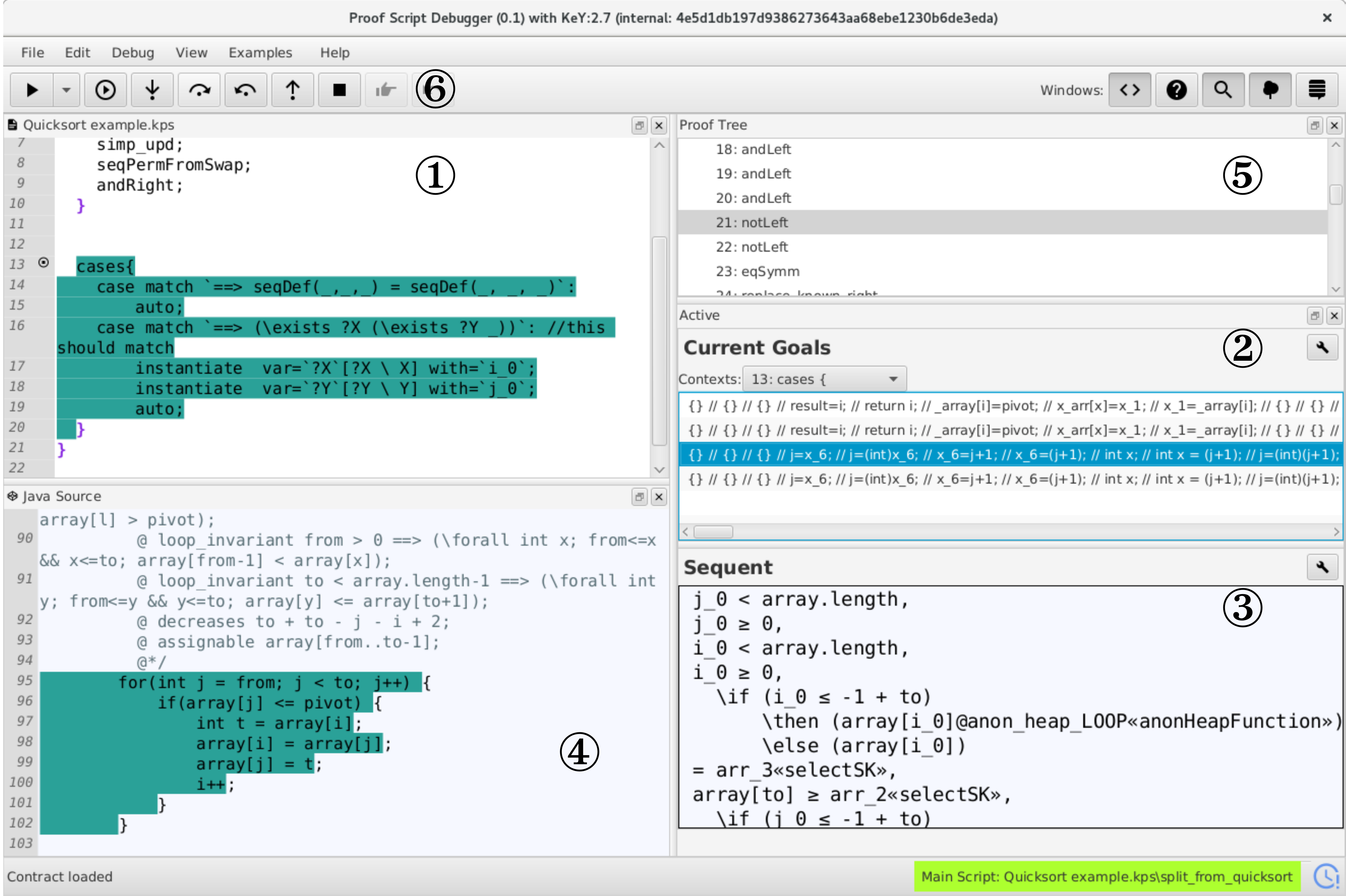}
  \caption{The user interface of \PSDBG}
  \label{fig:ui}
\end{figure}

\myparagraph{Visualization}
%
%
Like software debuggers, \PSDBG{} offers different views on the proof states
as shown in Fig.~\ref{fig:ui}:
\ding{172}~The source code of the proof script, with the next
command to be executed being highlighted. \ding{173}~A list of the current
proof goals and the; the currently selected
proof goal is highlighted. This window pane allows different representations
of the goals to be used, e.g., branching labels which are introduced by the underlying
verification system to identify certain proof branches like induction base, step and use case.
\ding{174}~Below, the selected proof goal is shown in full textual
representation; this view supports the application of rules on
selected terms in the interactive mode.  \ding{175}~The lower left pane shows
the source code of the Java program being verified.
The highlighted lines are the executed Java statements corresponding to the selected
proof goal. \ding{176}~The proof tree, i.e., the explicit proof
object constructed by~\KeY is displayed. Note that, only a small portion of the proof 
tree can be seen, showing the beginning of
the proof, where no branching has occurred yet.
\ding{177} The toolbar contains the buttons that are used
for stepping through the proof script's execution.
Not shown in the screenshot of Figure~\ref{fig:ui} is the editor for writing
and evaluating match expressions and the window with proof command documentation.
Note that not all
views are open all the time -- rather the user may choose which views to see.

%

\myparagraph{Breakpoints and stepping}
For the analysis of proof script executions, \PSDBG{} allows to set
breakpoints and to use stepping functionalities.
The tool supports line breakpoints with and without a boolean condition.
Script execution pauses when a breakpoint is reached and -- in case a
condition is provided -- if moreover
the state reached satisfies the breakpoint's condition.

When script execution is paused, the user can use the stepping functionalities
(by using the respective buttons in the toolbar).

For stepping, statements can either be compound
(blocks or prover strategies) or atomic (e.g.,
single rule applications or variable assignments).
The functions \emph{step into} and \emph{step over} have the
usual behavior known from software debugging: \emph{step over} executes until the end of the compound
statement, while \emph{step into} allows the user to inspect the execution of
the constituents of a compound statements in more detail (e.g., if invoked
before a block or a call to a sub-script). If \emph{step into} is invoked for
a native command of the underlying proof system,
there are two possibilities: if the proof command is a prover strategy,
the user is presented with the partial proof tree that corresponds to the
execution of the that strategy. Stepping into a single (atomic) calculus rule
behaves like \emph{step over}.

In addition to \emph{step over} and \emph{step into}, two more stepping
functions are available to inspect script execution in reverse: \emph{step over reverse} and
\emph{step into reverse}. These allow the user
to inspect proofs from end to start.
This reverse inspection of proof states is possible due to
the (partial) explicit proof object provided by the underlying verification
system.

\myparagraph{Interactive manipulation of proof goals}
When the execution of a script is completed  with some open proof goals remaining,
the user has the possibility to interactively manipulate these open goals
(e.g., using point-and-click interaction provided by the underlying verification system).
Our tool allows to make these user interactions persistent automatically by
recording and appending them to
the end of the proof script upon leaving the interactive mode.
%
%
%

\section{System Architecture}
\label{sec:architecture}
\begin{wrapfigure}[12]{r}{0.4\textwidth}
  \centering
  \includegraphics[width=0.4\textwidth]{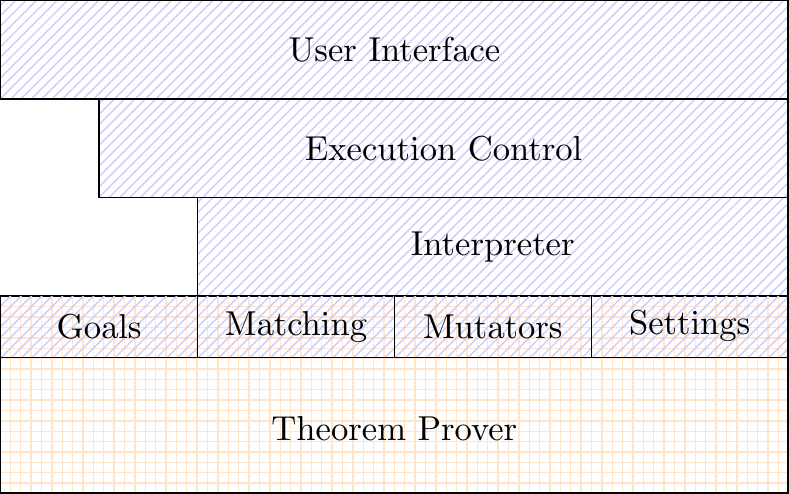}
  \caption{Block diagram of the architecture. The blue hatched parts are
    provided by our tool.}
  \label{fig:architecture}
\end{wrapfigure}

\PSDBG{} consists of three main components (Fig.~\ref{fig:architecture}),
which are built on top of an underlying theorem prover.
(1)~The user interface needs direct access to the theorem prover to allow
interactive execution of goal mutators and to extract  information for
visualization, e.g., the executed Java source code lines.
Integration of new additional views and state projection is supported by the use
of a docking framework and full access to the underlying stack.
(2)~Execution control is a layer that provides the debugging logic to the~UI,
e.g., stepping, state tracing and breakpoints.
(3)~The interpreter is the heart of the architecture. It executes the proof script
and performs the calls to the goal mutators of the theorem prover.

The debugging logic is separated from the interpreter core and the user
interface. The UI is in parts dependent on the underlying theorem prover, i.e.,
the shape of the goals and the prover's user interaction style (in \KeY\ the
goals are sequents, and \KeY\ uses a point-and-click style for interaction).

\setlength\intextsep{0pt}

For the adaption to a different theorem prover, the interpreter provides
well-defined extension points---so the execution control and interpreter core
are independent to the kind of proof goals.
The extensions points are the handler of goal mutators, and special variables
(prover settings) and the evaluation of matchings against verification
conditions.
The matching mechanism supports a special language for pattern matching of proof
goals. The current pattern language is optimized for \KeY's sequents and needs
to be adapted when using other types of proof goals.


\section{Future Work}
%


For future work, we will explore the usability of \PSDBG and \KPS on larger and more
complex verification tasks, where script modularization becomes necessary.
We plan to better visualise relations between different views, e.g.,
showing the relation between the proof script and the program to be verified.
Also, better support for proof exploration is planned, so that less
manual effort is required.


\medskip\noindent
\paragraph{Acknowledgements.}
Special thanks go to An Thuy Tien Luong who provided
valuable comments concerning the usage of \PSDBG and the proof scripting language.


\bibliographystyle{splncs}
\bibliography{refs}

\clearpage

\appendix
\section{QuickSort.java}
\label{sec:quicksortJava}

\begin{lstlisting}[language={[JML]Java}]
/**
 * This example formalizes and verifies the wellknown quicksort
 * algorithm for int-arrays algorithm.  It shows that the array
 * is sorted in  the end and that it contains  a permutation of
 * the original input.
 *
 * The   proofs   for   the  main   method   sort(int[])   runs
 * automatically   while   the   other  two   methods   require
 * interaction.  You   can  load   the  files   "sort.key"  and
 * "split.key"  from the  example's  directory  to execute  the
 * according proof scripts.
 *
 * The permutation property requires some interaction: The idea
 * is that the only actual modification on the array are swaps
 * within the "split" method. The sort method body contains
 * three method invocations which each maintain the permutation
 * property. By a repeated appeal to the transitivity of the
 * permutation property, the entire algorithm can be proved to
 * only permute the array.
 *
 * To establish  monotonicity, the key  is to specify  that the
 * currently  handled block  contains  only  numbers which  are
 * between   the    two   pivot   values    array[from-1]   and
 * array[to]. The first  and last block are exempt  from one of
 * these  conditions  since  they have  only  one  neighbouring
 * block.
 *
 * The  example has  been  added  to show  the  power of  proof
 * scripts.
 *
 * @author Mattias Ulbrich, 2015
 */

class Quicksort {

    /*@ public normal_behaviour
      @  ensures \dl_seqPerm(\dl_array2seq(array),
      @                          \old(\dl_array2seq(array)));
      @  ensures (\forall int i; 0<=i && i<array.length-1;
      @                     array[i] <= array[i+1]);
      @  assignable array[*];
      @*/
    public void sort(int[] array) {
        if(array.length > 0) {
            sort(array, 0, array.length-1);
        }
    }

    /*@ public normal_behaviour
      @  requires 0 <= from;
      @  requires to < array.length;
      @  requires from > 0 ==> (\forall int x; from<=x &&
      @                            x<=to; array[x] > array[from-1]);
      @  requires to < array.length-1 ==>
      @            (\forall int x; from<=x && x<=to; array[x] <= array[to+1]);
      @  ensures \dl_seqPerm(\dl_array2seq(array), \old(\dl_array2seq(array)));
      @  ensures (\forall int i; from<=i && i<to; array[i] <= array[i+1]);
      @  ensures from > 0 ==>
      @           (\forall int x; from<=x && x<=to; array[x] > array[from-1]);
      @  ensures to < array.length-1 ==>
      @                (\forall int x; from<=x && x<=to; array[x] <= array[to+1]);
      @  assignable array[from..to];
      @  measured_by to - from + 1;
      @*/
    private void sort(int[] array, int from, int to) {
        if(from < to) {
            int splitPoint = split(array, from, to);
            sort(array, from, splitPoint-1);
            sort(array, splitPoint+1, to);
        }
    }

    /*@ public normal_behaviour
      @  requires 0 <= from && from < to && to <= array.length-1;
      @  requires from > 0 ==> (\forall int x; from<=x && x<=to;
      @                                array[from-1] < array[x]);
      @  requires to < array.length-1 ==> (\forall int y; from<=y && y<=to;
      @                                            array[y] <= array[to+1]);
      @  ensures \dl_seqPerm(\dl_array2seq(array), \old(\dl_array2seq(array)));
      @  ensures from <= \result && \result <= to;
      @  ensures (\forall int m; from <= m && m <= \result;
      @                     array[m] <= array[\result]);
      @  ensures (\forall int n; \result < n && n <= to;
      @                     array[n] > array[\result]);
      @  ensures from > 0 ==> (\forall int x; from<=x && x<=to;
      @                     array[from-1] < array[x]);
      @  ensures to < array.length-1 ==> (\forall int y; from<=y && y<=to;
      @                     array[y] <= array[to+1]);
      @  assignable array[from..to];
      @*/
    private int split(int[] array, int from, int to) {

        int i = from;
        int pivot = array[to];

        /*@
          @ loop_invariant from <= i && i <= j;
          @ loop_invariant from <= j && j <= to;
          @ loop_invariant \dl_seqPerm(\dl_array2seq(array),
          @                                \old(\dl_array2seq(array)));
          @ loop_invariant (\forall int k; from <= k && k < i; array[k] <= pivot);
          @ loop_invariant (\forall int l; i <= l && l < j; array[l] > pivot);
          @ loop_invariant from > 0 ==>
          @       (\forall int x; from<=x && x<=to;  array[from-1] < array[x]);
          @ loop_invariant to < array.length-1 ==>
          @       (\forall int y; from<=y && y<=to; array[y] <= array[to+1]);
          @ decreases to + to - j - i + 2;
          @ assignable array[from..to-1];
          @*/
        for(int j = from; j < to; j++) {
            if(array[j] <= pivot) {
                int t = array[i];
                array[i] = array[j];
                array[j] = t;
                i++;
            }
        }

        array[to] = array[i];
        array[i] = pivot;

        return i;

    }
}
\end{lstlisting}

\end{document}